# How and where to look for tRNAs in Metazoan mitochondrial genomes, and what you might find when you get there

David A. Morrison

Section for Parasitology (SWEPAR)
Department of Biomedical Sciences and Veterinary Public Health
Swedish University of Agricultural Sciences, 751 89 Uppsala, Sweden
Email: David.Morrison@bvf.slu.se

**Abstract**

The ability to locate and annotate mitochondrial genes is an important practical issue, given the rapidly increasing number of mitogenomes appearing in the public databases. Unfortunately, tRNA genes in Metazoan mitochondria have proved to be problematic because they often vary in number (genes missing or duplicated) and also in the secondary structure of the transcribed tRNAs (T or D arms missing). I have performed a series of comparative analyses of the tRNA genes of a broad range of Metazoan mitogenomes in order to address this issue. I conclude that no single computer program is necessarily capable of finding all of the tRNA genes in any given mitogenome, and that use of both the ARWEN and DOGMA programs is sometimes necessary because they produce complementary false negatives. There are apparently a very large number of erroneous annotations in the databased mitogenome sequences, including missed genes, wrongly annotated locations, false complements, and inconsistent criteria for assigning the 5' and 3' boundaries; and I have listed many of these. The extent of overlap between genes is often greatly exaggerated due to inconsistent annotations, although notable overlaps involving tRNAs are apparently real. Finally, three novel hypotheses were examined and found to have support from the comparative analyses: (1) some organisms have mitogenomic locations that simultaneously code for multiple tRNAs; (2) some organisms have mitogenomic locations that simultaneously code for tRNAs and proteins (but not rRNAs); and (3) one group of nematodes has several genes that code for tRNAs lacking both the D and T arms.

# 1 Introduction

Transfer RNAs (tRNAs) are key molecules in protein biosynthesis in all living organisms, occurring as a single cytosolic set in prokaryotes but with additional sets for each organelle in eukaryotes [1]. Metazoan mitochondria-encoded tRNAs differ from their nuclear-encoded counterparts in that there are fewer of them and they are much more variable in structure, in terms of differences both between tRNA types and between taxonomic groups [2, 3]. There are usually only 22 (instead of 61) different tRNAs in each mitochondrial genome (mitogenome), but there may be more or many fewer; and the standard cloverleaf-shaped secondary structure may lack one of the cloverleaf arms [2, 4]. This variability means that we currently know less about how they function in the mitochondrion compared to the more standardized nuclear tRNAs, or what biochemical significance the many variants may have [5].

At a more practical level, the variability means that it is sometimes very difficult to locate all of the genes within the mitogenome and to precisely annotate their 5' and 3' boundaries [6]. A similar problem applies to protein-coding genes, which have variable start codons and truncated stop codons, and also to rRNA-coding genes, which are poorly conserved at their 5' and 3' ends. However, here I am mainly concerned with tRNA-coding genes.

In this paper, I report the results of comparative analyses of the tRNA genes of a broad range of Metazoan mitogenomes. The diversity of topics that I cover include: (1) how best to search for the full complement of tRNA genes; (2) the types of annotation inconsistencies and errors that occur when contradictory strategies and philosophies are adopted; (3) how much overlap there is between the coding regions of tRNAs, proteins and rRNAs; (4) the possibility that there might be genomic locations that simultaneously code for multiple tRNAs; (5) the possibility of genomic locations that simultaneously code for proteins and tRNAs; and (6) the possibility that there might be genes coding for tRNAs that lack both D and T arms.

It is important to note at the outset that I am using consistency of the gene annotations across closely related species as the arbitrator of correctness of those annotations, rather than the plausibility of any molecular mechanisms that might be implied, in the absence of any information regarding the sequences of the molecules coded by the genes. If this is an inappropriate criterion for any particular situation, then my comparative analyses may not be successful. Furthermore, my first intention is solely to highlight and evaluate the nature and extent of problems with annotation of mitochondrial genomes. I do not offer any explicit solutions to these problems other than the sort of detailed comparative analyses that I present in this paper. My second intention is solely to propose some rather bold hypotheses regarding mitochondrial tRNAs, and to evaluate the bioinformatic evidence in their favor. I do not present any experimental evidence either for or against them.

# 2 Materials and Methods

The source collections of the tRNA genes considered here are from an arbitrary selection of taxonomic groups where unusual or unexpected tRNA features have been referenced in the literature. I did not attempt to comprehensively cover the known mitochondrial tRNAs, and thus it is not known how representative my collection is (e.g. there are some poorly annotated genomes included). The collections included did sample a wide selection of Metazoan groups, with the notable exception of the Mammalia, including: Porifera, Cnidaria, Echinodermata, Mollusca, Nematoda, Chelicerata (two data sets), Crustacea, Insecta (two data sets) and Lepidosauria. The mitogenomes and alignments are listed in Appendix 1 (Supplementary Online Material).

The tRNA gene sequences were downloaded from the OGRe database [7] (http://drake.physics.mcmaster.ca/ogre/) where possible (these were usually the NCBI RefSeq genomes), and otherwise from the GenBank database. In cases where alignment discrepancies were identified (as described below), both sources were used, as well as any sequences shown in the original publications (e.g. in tRNA or rRNA secondary structure diagrams).

The locations of the tRNA genes within the mitochondrial genomes could be determined in one or more of five possible ways. First, most of the gene locations were identified in the original publications and/or in the associated GenBank sequence annotations. Second, the tRNAscan-SE v. 1.21 program [8] was employed, using the web interface (http://lowelab.ucsc.edu/tRNAscan-SE/) and the "Nematode Mito" settings for the COVE program. Third, the optimized implementation of the COVE program provided in the DOGMA package [9] was employed, using the web interface (http://dogma.ccbb.utexas.edu/) and a "COVE threshold" of 5 (very low). Fourth, the ARWEN v. 1.2 program [10] was employed, using the web interface (http://130.235.46.10/ARWEN/) with the "mtmam" option switched off. Fifth, the mitogenomes were searched manually using the "Find" facility of a word processor, searching both strands for motifs that matched the anticodon stem/loop sequence. Other possible tRNA-search programs also exist, such as RNAMotif [11] and tRNAfinder [12], but they were not examined here.

Annotation of the tRNA genes was then performed in the context of a multiple alignment of all of the available sequences for each defined taxonomic group, which is recognized as being the most powerful form of comparative analysis [3, 13,14]. None of the current automated annotation tools explicitly provide this facility (most work pairwise only, such as when relying on BLAST matches), so all of my alignments were constructed manually.

If these alignments are "correct" in the comparative sense then they should have two characteristics [3]. First, there should be identifiable nucleotide motifs related to the current hypotheses about phylogenetic relationships among the sequences (or as reflected in their taxonomy). These motifs will appear as vertical patterns in the alignment (in its standard orientation). Second, there should be consistency of pairing of the nucleotides in accord with the current hypotheses of the secondary structure of mitochondrial tRNAs. That is, there will also be horizontal patterns in the alignment, where both nucleotides of each stem-pair should be aligned across all of the sequences (this is called structural consistency). The hypotheses of phylogenetic relationship were taken from the original publications and/or the NCBI Taxonomy. The tRNA structures were initially compiled from the secondary structure diagrams provided in the original publications, and were then fitted to a standardized mitochondrial tRNA model (Fig. 1). Note that this model is intentionally simple, and does not take into account either possible conserved nucleotides or tertiary interactions. The objective here is to evaluate tRNA prediction and annotation as it is currently practiced, and this model suffices for this purpose because this extra information has almost never been used (and is generally not available).

If both of these characteristics were present in all of the sequences in an alignment then the alignment was considered to be unproblematic. Any sequences that deviated in either characteristic were manually realigned to try to resolve the discrepancies. If this was unsuccessful, then three possible explanations were considered for each offending sequence: (1) the annotation of the sequence was wrong; (2) there was an issue with the nucleotide sequence itself; and (3) the hypothesized secondary structure of the tRNA was wrong. Examples of (1) included mis-specified 3' or 5' ends, and incorrect location of the gene within the genome. Examples of (2) included missing/extra nucleotides, possibly the result of sequencing errors. Examples of (3) included variation in stem or loop lengths, changed structural roles, and presence/absence of stems (notably the D or T arm).

The alignments presented here are the final result of the comparative analysis, each one explicitly identifying which of the three possible errors I consider to be most likely for each discrepancy identified.

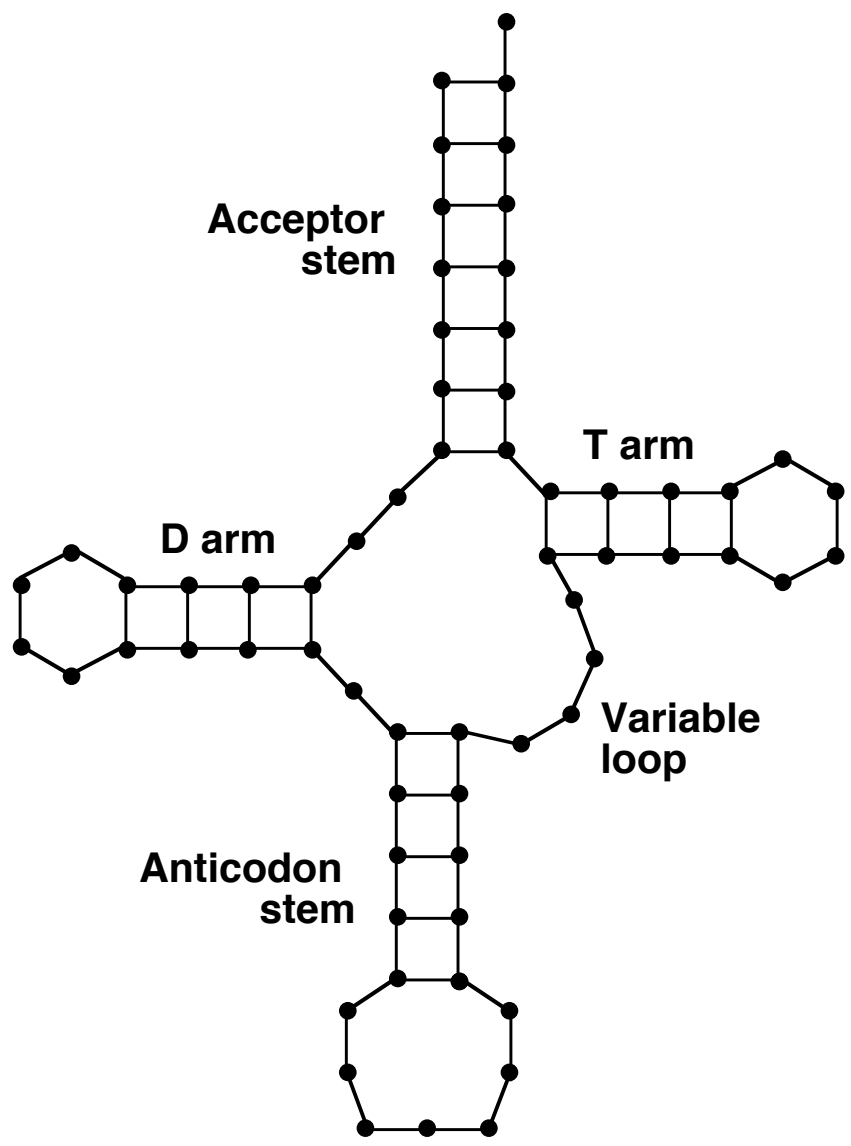

**Fig. 1.** A schematic representation of the secondary-structure model used for Metazoan mitochondrial tRNAs. Each nucleotide is represented by a dot, with the lines representing bonds. The sequence starts at the top left, with the final discriminator nucleotide at the top right. The D arm consists of the DHU stem and loop, and the T arm consists of the TΨC stem and loop. The Anticodon stem supports a loop with the 3-base anticodon. The lengths of the D arm and particularly the T arm are apparently quite variable among species, and may thus be longer or shorter than shown here.

# 3 Results and Discussion

## *3.1 Detecting tRNAs*

The most commonly cited computer program for locating tRNA genes in mitochondrial genomes was tRNAscan-SE. However, this was originally optimized for finding nuclear tRNAs, where it can perform well [1], but it appears often to perform poorly for Metazoan mitochondrial genes. This is so even when it is set to the "Nematode mito" setting, which allows for missing D or T arms, as it finds many fewer than half of the tRNA genes in most nematode mitogenomes. Similar poor performance has been observed in large-scale comparative studies by Wyman and Boore [15], Laslett and Canbäck [10] and Abe et al. [16], and at a smaller scale by Masta and Boore [17].

These poor results occur because the default COVE cut-off score is 20, and few Metazoan mitochondrial tRNAs score above this level. As a single example, tRNAscan-SE found 78 potential tRNA genes in the *Mesobuthus gibbosus* (Arachnida; Scorpiones) mitogenome, 20 of which appeared to be correct but only 4 of which had scores >20 (Fig. 2). One solution to this problem used by a number of authors was lowering the cut-off score (e.g. [17, 18] used 0.1). However, this can lead to ridiculous effects, such as for *M. gibbosus* where a score of 0.38 was used to annotate a possible tRNA-asp that matched the secondary structure no better than would a random sequence [19].

When identifying possible tRNA genes, the ARWEN progam produced more true positives but also more false positives than did tRNAscan-SE. These extra true positives mean that a number of published reports that some tRNA genes are missing from certain mitogenomes are erroneous, a point also noted by Perseke et al. [20]. Examples of genes found by ARWEN but not tRNAscan-SE included (the details are listed in Appendix 2): (i) tRNA-ser(AGN) in an unannotated space for *Aleurodicus dugesii* (Arthropoda; Insecta); (ii) tRNA-leu(CUN) at the end of what was annotated as the control region of *Phalangium opilio* (Arthropoda; Opiliones) (NB: [17] mistakenly claimed that it is tRNA-

leu(UUR) that is missing); (iii) tRNA-asp and tRNA-glu in unannotated spaces for *Aphrocallistes vastus* (Porifera), and also tRNA-leu(UUR) at the end of what was annotated as the control region; (iv) tRNA-asn in an unannotated space for *Tetraleurodes acaciae* (Arthropoda; Insecta), and tRNA-ile exactly where it would be predicted from comparison with related species; (v) tRNA-ile for *Neomaskellia andropogonis* (Arthropoda; Insecta) exactly where it would be predicted from comparison with related species; (vi) similarly for tRNA-ser(AGN) of *Schizaphis graminum* (Arthropoda; Insecta) (NB: this gene is currently mis-annotated in the NCBI RefSeq); and (vii) tRNA-ser(UCN) for *Xiphinema americanum* (Nematoda), which mostly overlaps with the NAD5 gene.

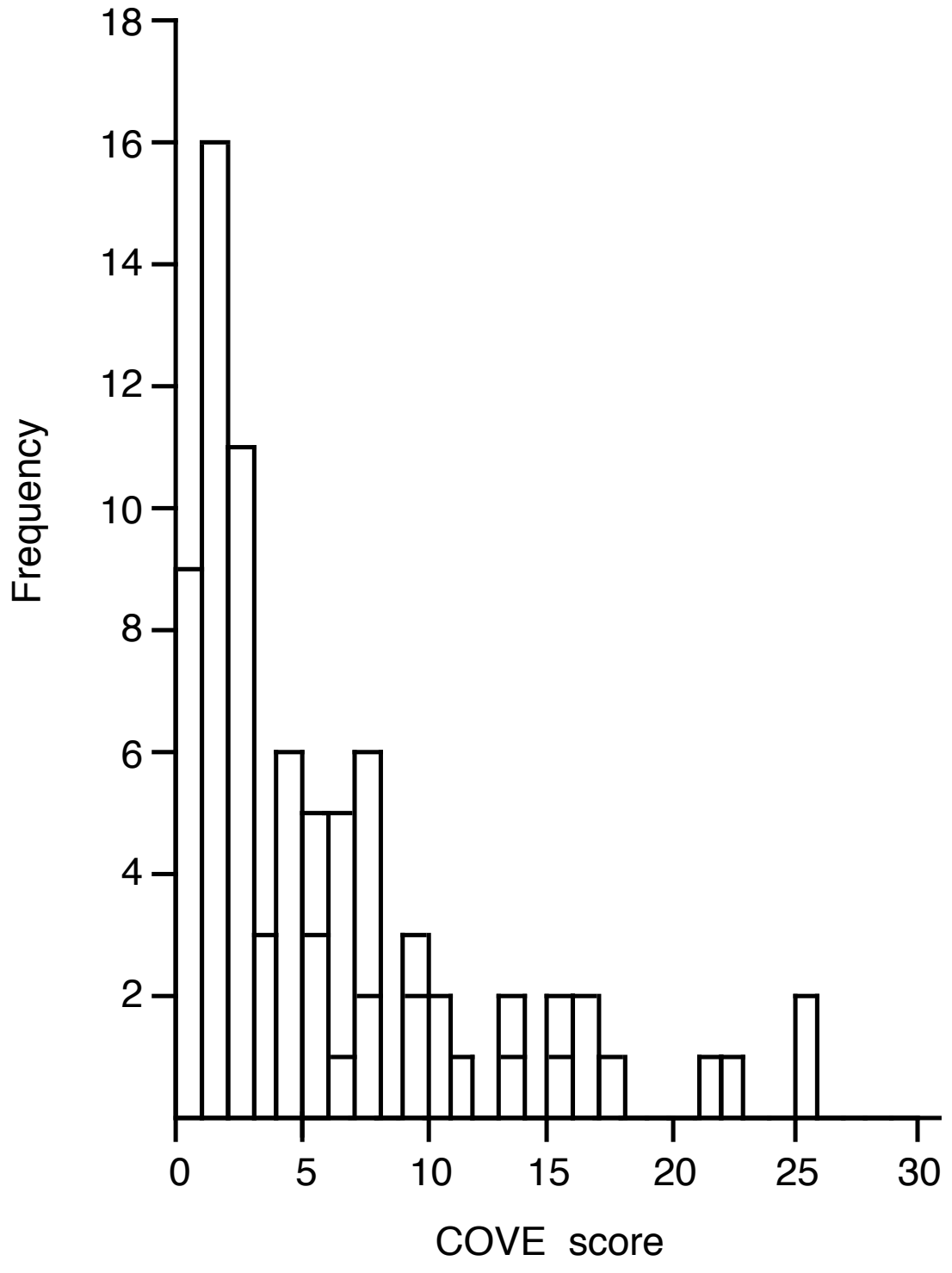

**Fig. 2.** Frequency histogram of the COVE scores obtained from analyzing the *Mesobuthus gibbosus* (Arachnida; Scorpiones) mitochondrial genome with the tRNAscan-SE program. The solid bars represent the hypothesized true positives and the open bars the false positives.

A slightly more complicated example involved the tRNA-val gene of *Ligia oceanica* (Arthropoda; Crustacea). The originally annotated sequence produced a tRNA secondary structure that matched my model rather poorly, whereas the location identified by ARWEN produced a much better fit. However, this alternate location lies within what has been annotated as the l-rRNA gene. So, I constructed an alignment of the 21 Malacostraca l-rRNA sequences with reference to the secondary structure diagram for the *Artemia salina* (Crustacea: Branchiopoda) sequence from the CRW web site [21]. This showed that the 3' end of the hypothesized *L. oceanica* tRNA-val sequence is just near the first l-rRNA stem, and thus it may not overlap at all with the l-rRNA gene (i.e. the most likely l-rRNA 5' end is immediately after the tRNA-val).

A slightly different example involved the tRNA-trp gene of *Savalia savaglia* (Cnidaria). The sequence at the location identified by ARWEN was clearly related to the gene sequences of the other species in terms of motifs, but it had several stem pairs missing and had 2 nt extra in the Variable loop. It may thus be a pseudogene.

In spite of the increased number of true and false positives, ARWEN still produced false negatives. These negatives could sometimes be found by the COVE program with the optimized settings implemented in the DOGMA package, by using a very low threshold score. An example of a gene found by DOGMA but not by ARWEN was tRNA-thr in an unannotated space for *Aphrocallistes vastus* (Porifera) (the details are listed in Appendix 2).

A more complicated example involved the tRNA-arg for *Neomaskellia andropogonis* (Arthropoda; Insecta). The location identified by DOGMA lies within what has been annotated as the COX2 gene. So, I constructed an alignment of the 11 Hemiptera COX2 sequences, which showed that the *N. andropogonis* sequence would produce a protein that is 17–18 amino acids longer than those of its relatives. Re-annotating the 3' end of the COX2 gene (see below) removes any overlap with the hypothesized tRNA-arg gene sequence. Unfortunately, the tRNA sequence is not very similar to those of the related species.

A slightly different example involved the tRNA-gln gene of *Aleurodicus dugesii* (Arthropoda; Insecta). The location identified by DOGMA lay within what has been annotated as the NAD5 gene. Since re-annotation does not resolve the discrepancy, I discuss this situation in more detail in a later section.

Unfortunately, DOGMA itself often missed tRNA gene locations suggested by ARWEN. In particular, ARWEN and DOGMA often missed potential tRNA genes that are the reverse-complement of each other (discussed in more detail below). Thus, no single program is adequate for locating potential tRNA genes (cf. [16]). In addition, DOGMA produced many false positives. The problem seems to be that the COVE scores are based on the number of stem pairs, so that long T or D arms can produce high scores even if the Anticodon or Acceptor stems is badly mis-paired.

Perhaps even more unfortunately, some of the potential tRNA locations missed by all three computer programs could be found by searching manually for the specified anticodon motif. Indeed, the majority of the original authors reported finding some of their annotated genes manually, although most of them did not use either ARWEN or DOGMA. An example of a gene found in this way was a potential tRNA-arg gene of *Ligia oceanica* (Arthropoda; Crustacea) in what was annotated as the control region. The implied tRNA secondary structure is shown in Fig. 3c; and some of the complications with this particular example are discussed in more detail below.

Sometimes, the tRNA genes were not found by any of the programs because the genes remain unsequenced. A particularly egregious example was provided by the tRNA-glu and tRNA-thr genes of *Pisaster ochraceus* (Echinodermata). The current NCBI RefSeq (NC_004610; 14,837 nt) is annotated as being a complete circular mitogenome, in spite of the fact that it was taken unchanged from accession X55514.2 which is explicitly annotated as an incomplete linear genomic sequence. (Note that X55514.2 is itself an updated version of X55514.1, adding 6,809 nt to the original 8,028 nt of sequence.) The tRNA-glu and tRNA-thr gene sequences are not included in X55514 but instead are in accession M25320 [22, 23]. These two accessions overlap by 2 nt, and so they could easily be combined. The combined sequence (15,490 nt) would cover all of the mitogenome except for the middle c.710–720 nt of the s-rRNA. This unnecessary confusion has affected the work of at least one group of researchers [20].

Finally, sometimes tRNA genes were not found by any of the programs because the genes really are absent from the mitogenome. Examples included: the tRNA-asp of the Buthidae (Arachnida; Scorpiones) (see below); and the Cnidaria, which appear to have *only* the tRNA-met and tRNA-trp genes [24].

No computer program currently uses the strategy of comparative analysis to identify mitochondrial tRNA genes (or any other type of gene, for that matter), although the DOGMA package and the procedure of Perseke et al. [20] utilize some of the ideas. It is not immediately obvious how to develop such a program, although the approach of Lagesen et al. [25] offers some promise. The idea here would be to use comparative analysis to create a reliable alignment for each tRNA and each relevant taxonomic group (as I have done here), and then to train a hidden markov model (HMM) for each of these alignments. The RNAmmer program could then be used in conjunction with this library of HMMs to annotate new mitogenomes.

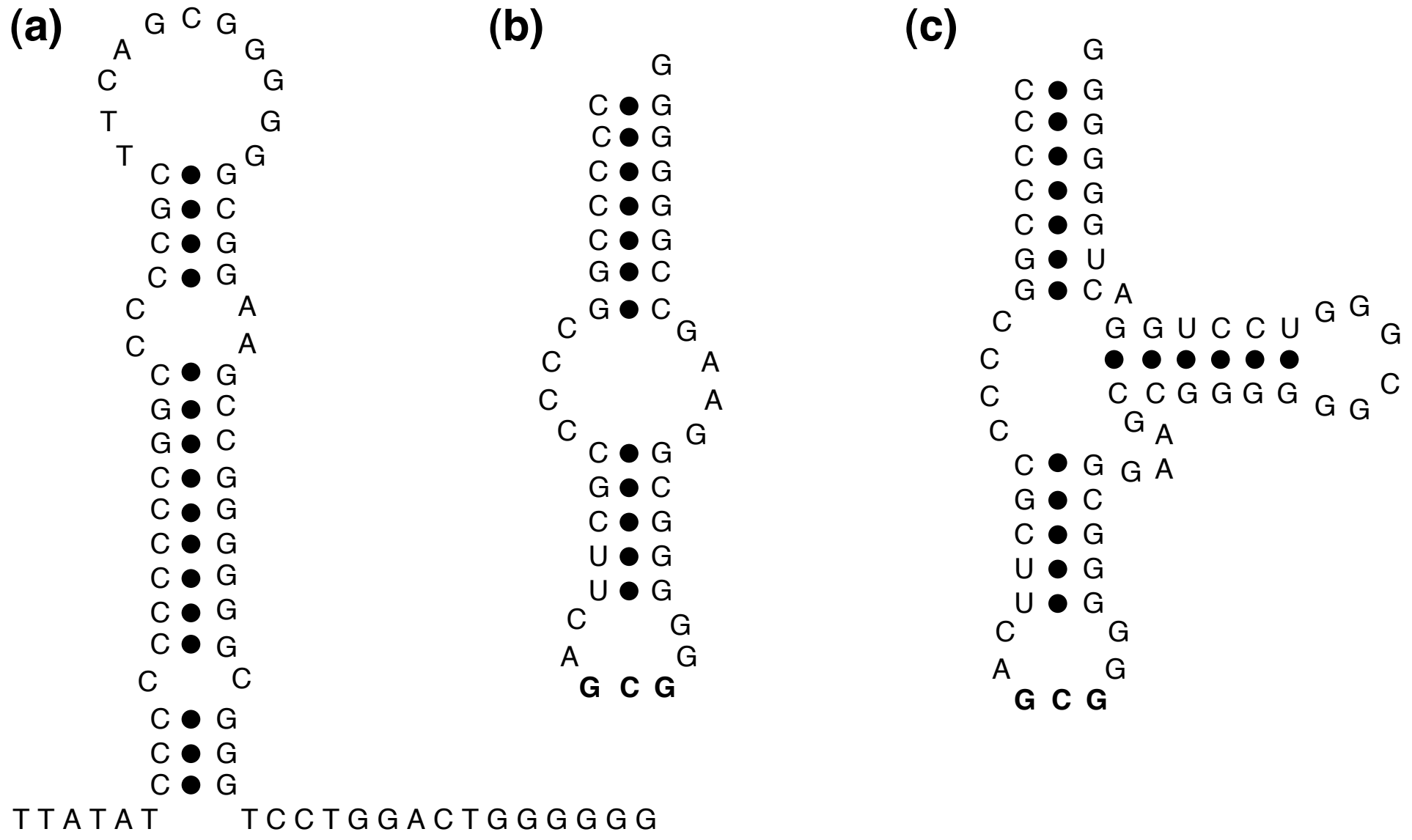

**Fig. 3.** Part of the mitochondrial genome of *Ligia oceanica* (Arthropoda; Crustacea) that could be annotated in three different ways: (a) as a stem-loop in the control region (its current annotation); (b) as the gene for a D-armless and T-armless tRNA-arg; and (c) as the gene for a D-armless tRNA-arg. Dots indicate paired nucleotides and boldface highlights the anticodon.

## *3.2 Annotation errors*

Most of the alignment discrepancies among the tRNA genes could be resolved by re-annotating one or more of the sequences. It thus seems that an important part of locating tRNA genes is getting the boundaries correct. I will therefore discuss this topic here with reference to all types of genes (including protein- and rRNA-coding), giving a few examples to indicate the diversity of problems. All of the apparent annotation errors are documented in Appendix 2.

If annotations are inconsistent when compared across a taxonomic group, then presumably one or more of the annotations is incorrect. This is an inevitable but unfortunate consequence of having different people annotating different genomes using different strategies and with different philosophies or expectations. My approach here was to adopt a consistent set of annotation decisions across all of the taxonomic groups. It is important to emphasize my use of consistency as the arbiter of correctness [25]. Thus, even if my annotations are incorrect in some way they are at least consistently wrong, and can thus be easily corrected. A similar approach was adopted by Lavrov et al. [26], who also suggested some explicit rules for determining mitochondrial gene boundaries.

The most obvious way in which different philosophies or expectations affect annotations is with regard to anticipated overlap between genes, and this is discussed in more detail in the next section. In this section, I will restrict myself to more practical sources of inconsistency.

Locating the ends of almost all mitochondrial genes can be a problem. For protein-coding genes, there is no standard start codon and there may be truncated stop codons, and the ends of the proteins themselves are often not highly conserved. As one example of a recurrent problem, which was also noted by Lavrov et al. [26], the COX2 gene of *Neomaskellia andropogonis* (Arthropoda; Insecta) was annotated as using a TAA as the stop codon, which made it much longer than the genes of related species (see above). Re-annotating the 3' end to use a truncated T codon, instead, produces a gene of comparable length.

When locating the ends of tRNA genes, most authors seemed to rely on the fit to some tRNA structural model, and this model often varied greatly between authors. In most of the alignments the standard model that I used required at least one sequence to have its end(s) adjusted, usually in some minor way. There are 32 tRNA sequences re-annotated in this way listed in Appendix 2.

Perhaps the most obvious problem in the mitochondrial genomes was locating the ends of rRNA genes, since neither end of either rRNA is highly conserved. This has long been a problem even for nuclear rRNA genes [25, 27]. The most common strategy was to simply annotate the ends with reference to the ends of the preceding and succeeding genes. This will likely over-estimate the true length of the rRNA genes. For example, the l-rRNA alignment for the Malacostraca (referred to above) suggested that the 5' end of all of the sequences except for *Cherax destructor* may have mis-annotations of 30–80 nt, and that the 3' end of all of the sequences except for *Geothelphusa dehaani* may have mis-annotations of 10–40 nt.

Another problem encountered was the annotated orientation of the genes, which was also noted by Lagesen et al. [25] and Perseke et al. [20]. For example, the tRNA-ser(AGN) of *Albinaria caerulea* (Mollusca) should not be the complement in the annotation, nor should the s-rRNA gene in the RefSeq sequence of *Varroa destructor* (Arthropoda; Chelicerata). The alternative mitogenome of *Varroa destructor* (AY163547) has seven genes that were shown (correctly) as complemented in the publication but not in the database annotations.

Finally, some genes were annotated in the wrong locations, either in the database entry or in the original publication. For example, the tRNA-ser(AGN) gene of *Aphrocallistes vastus* (Porifera) was annotated in the GenBank sequence and Table 1 of the paper as being tRNA-arg, but Figure 1 of the paper annotated it correctly. The tRNA-gln gene of *Varroa destructor* (AY163547) was annotated as c(12127..12861), which clearly cannot be correct (based on length alone!). The more likely annotation c(12061..12127) involves swapping the two numbers plus a typographical error. The l-rRNA and s-rRNA genes of *Marsupenaeus japonicus* (Arthropoda; Crustacea) had clearly had their labels interchanged. The tRNA-val gene of *Cherax destructor* (Arthropoda; Crustacea) simply did not match those of related species, and a well-matched alternative could be found at the end of what was annotated as the control region.

It is clear that some sort of consistent strategy for the annotation of mitochondrial genomes is needed, otherwise the sorts of inconsistencies reported here will continue to proliferate. A number of authors have provided detailed descriptions of protocols that, if adopted, would at least lead to consistency, including Lavrov et al. [26], Masta and Boore [17] and Perseke et al. [20]. It may even be feasible to develop a formal standard, along the lines of those promoted by the MIBBI project (Minimum Information for Biological or Biomedical Investigations; http://www.mibbi.org/).

Given the widespread occurrence of annotation discrepancies, it is unfortunate that database accessions (including GenBank) cannot be annotated except by the original authors [28]. This means that third-party re-annotations will go unrecognized by future researchers unless they are published in the literature [20, 26] and the relevant publication happens to be consulted. This situation contrasts strongly with the tradition in the systematics community, where all herbarium/museum/culture specimens can have supplementary annotations added by experts without replacing the original annotation. All supplementary annotations for a specimen are freely available for perusal by being physically linked to the original annotation; and this system has worked well for several hundred years. The molecular-biology community could learn a valuable lesson from this arrangement.

Perhaps the most obvious suggestion is to create a supplementary database where researchers can submit annotations of the sequences associated with existing accession numbers, with a link added to the original DDBJ/EMBL/GenBank record. The alternative idea of a centralized quality-checked reference set of data seems to be a rather impractical compared to having supplementary annotations of the original records. Swiss-Prot is the best known attempt to do this, but it cannot be scaled-up to the requirements of nucleotide databases. NCBI's RefSeq database [29] is only a half-hearted attempt at this, as there is no actual quality control of the data, only consistency of presentation, and many of the records have blatant inconsistencies (some of which did not exist in the original record!). However, additions to the annotations database could be manually evaluated and incorporated into the NCBI RefSeq records.

So, while it may seem faintly ridiculous to have a database of annotations for another database, rather than having one high-quality database in the first place, there seem to be few alternatives that will be acceptable to the curators of the current databases. An alternative, but less practical, approach is to rely on quality control in the downstream databases, such as OGRe [7], tRNAdb [30], MetAMiGA [31], Mamit-tRNA [32] and PLMItRNA [33]. Propagation of downstream re-annotations back to the original sequence owners was apparently suggested for the Greengenes nucleotide database [34], but this remains as the only unimplemented part of their original strategy (http://greengenes.lbl.gov/).

### *3.3 Gene overlap*

When annotating their genomes, most authors first identified the protein-coding genes, and then searched for rRNA and tRNA genes only in the remaining unassigned regions (although [35] found the tRNAs first and then fitted the others into gaps). This is because overlap of coding regions is usually considered to be restricted to <5 nt, except for the well-documented overlap of ATP6-ATP8 and ND4-ND4L [4]. For example, Masta [36] and Masta and Boore [18] used apparently extensive overlap of tRNA genes as one of their main arguments for recognizing truncated tRNAs in spiders.

However, many long overlaps have been reported in the literature. My comparative studies showed that almost all of these long overlaps merely resulted from lack of consistency between sequence annotations, so that they could be resolved by adopting a consistent set of annotation decisions across all of the taxonomic groups. A few diverse examples are presented here; and the relevant alignments are shown in Appendix 3.

The tRNA-gly gene of *Oxya chinensis* (Arthropoda; Insecta) was annotated as having an overlap of 19 nt with the COX3 gene. However, this occurred because a TAG was used as the stop codon in the COX3 gene, whereas there is a truncated TA motif that aligns well with the stop codons in the sequences of all of the related species. The re-annotation would leave a gap of 3 nt between the genes. Similarly, the tRNA-gly of *Priapulus caudatus* (Priapulida) was annotated as having an overlap of 17 nt with the COX3 gene. Re-annotation to use an aligned T as a truncated stop codon (instead of TAA) would reduce the overlap to zero.

The tRNA-pro gene of *Enterobius vermicularis* (Nematoda) was annotated as having an overlap of 18 nt with the COX3 gene. Re-annotation of the tRNA-pro gene to match my model tRNA secondary structure would remove 1 nt of this overlap. Also, the COX3 gene was annotated to be longer than for all of the related species, because a TAG was used as the stop codon; and re-annotation to use a truncated T motif would remove another 5 nt. Finally, the protein-coding sequence is missing some strong motifs that occur in the other species, and so several equally likely alternative scenarios exist for the alignment. One of these would remove another 6 nt of the overlap while a different one would remove a further 6 nt. The latter scenario (i.e. removing 17 nt) would result in no overlap between the COX3 and tRNA-pro genes.

The tRNA-gly gene of *Euhadra herklotsi* (Mollusca) was annotated as having an overlap of 10 nt with the COX2 gene. However, there is an obvious +1 translational frameshift in the COX2 gene 99 nt upstream that affects the choice of stop codon. In the same alignment, *Cepaea nemoralis* (Mollusca) was annotated as having an overlap of 10 nt between the COX2 and tRNA-tyr genes. Similarly, there is a +1 translational frameshift in the COX2 gene 51 nt upstream that affects the choice of stop codon. The re-annotations would leave a gap of 1 and 0 nt between the genes, respectively. This alignment thus involved issues with the nucleotide sequences themselves rather than with the annotation (and I offer no explanation for the cause of the frameshifts).

*Aphrocallistes vastus* (Porifera) was annotated to have seven overlaps between genes > 5 nt long. Three of the overlaps (14, 19 and 28 nt) between tRNA and protein-coding genes were confirmed by my comparative analyses. Two of the remaining overlaps involved the NAD5 gene, which had its 3' end annotated to overlap with both the tRNA-cys (7 nt overlap) and tRNA-phe (71 nt) genes. This occured because the gene was annotated to end with a TAA stop codon, whereas the related species were annotated to use a truncated T motif as the stop codon, thus ending immediately before the tRNA-phe gene. Re-annotation of the *A. vastus* gene to use the aligned TA motif would leave a gap of 1 nt before the tRNA-phe gene, thus removing the need for either overlap.

Most of the gene overlaps reported in the literature were between tRNAs and proteins or between tRNAs and tRNAs. However, the *A. vastus* mitogenome was also annotated to have a 43 nt overlap between the COX3 and NAD2 genes. This occured because of an inconsistent choice of start codon for the NAD2 gene. Re-annotating this gene to use the ATG that aligns with the start codon of the related species would leave a gap of 5 nt.

The *A. vastus* mitogenome was also annotated to have a 43 nt overlap between the NAD2 and NAD5 genes. The comparative analysis was difficult because the other Hexactinellida mitogenomes (for *Iphiteon panicea* and *Sympagella nux*) were annotated as having undetermined stop codons. The alignment of the Demospongiae genes was unproblematic, but the alignment between the Hexactinellida and the Demospongiae was very ambiguous for the final third of the gene. An alignment of the translated amino acid sequences was thus prepared using the PROMALS program [37], via the web interface (http://prodata.swmed.edu/promals/promals.php), but to no avail. Consequently, it is likely that the location of the 3' end of the NAD2 gene of the Hexactinellida will only be determined by sequencing the transcribed proteins themselves.

As noted above, overlap involving rRNA genes was rarely reported, due to the way in which annotations were carried out. Two examples are discussed in the next section.

The only mitogenome that my comparative analyses reported as unambiguously having extensive overlap between several genes was that of *Epiperipatus biolleyi* (Onychophora). There were apparent inconsistencies in the original annotations of many of the genes throughout this genome (including the apparent absence of 9 tRNA genes), and so my comparative analysis ultimately involved re-annotating the entire mitogenome of this species. The procedures for producing the comparative alignments of the tRNA and rRNA genes were those described in previous sections, and the procedure for the protein-coding genes used the PROMALS program on the translated amino acid sequences (see above). Most of the comparisons used the mitogenomes of the nematode species, plus those of *Leptorhynchoides thecatus* (Acanthocephala), *Limulus polyphemus* (Arthropoda; Chelicerata) and *Taenia solium* (Platyhelminthes). The new annotations are listed in Appendix 4.

There were 22 gene boundaries that overlapped in this genome (Fig. 4a) after my comparative analysis, only five of which involved genes on opposite strands. All of the overlaps involved tRNAs, except for the ATP6 gene which overlapped with ATP8 by 4 nt and COX3 by 1 nt (all three of which were on the same strand). There were 14 unusually large (>10 nt) gene overlaps here (Fig. 4a) plus two instances in which one gene was completely overlapped by another (discussed further in the next section). All of these overlaps involved tRNA genes (Fig. 4b): 8 tRNA+tRNA, 7 tRNA+protein and 1 tRNA+rRNA.

Only 2/22 tRNAs had no gene overlap (tRNA-arg, tRNA-his), while 8/22 of the others had overlap at only one end; 10/22 had total overlaps of ≥50% of their length (Fig. 4b). There was more average overlap of tRNAs with proteins (24.6 nt; n=9) and rRNA (28 nt; n=1) than between tRNAs (17.3 nt; n=10). There was no difference in average overlap between the 5' (12.2 nt) and 3' (11.8 nt) ends of the tRNA genes. The two tRNAs with complete overlap (58 and 60 nt) were not the shortest ones, as tRNA-arg had 55 nt (with no overlap), but all of the others had lengths of 59–71 nt.

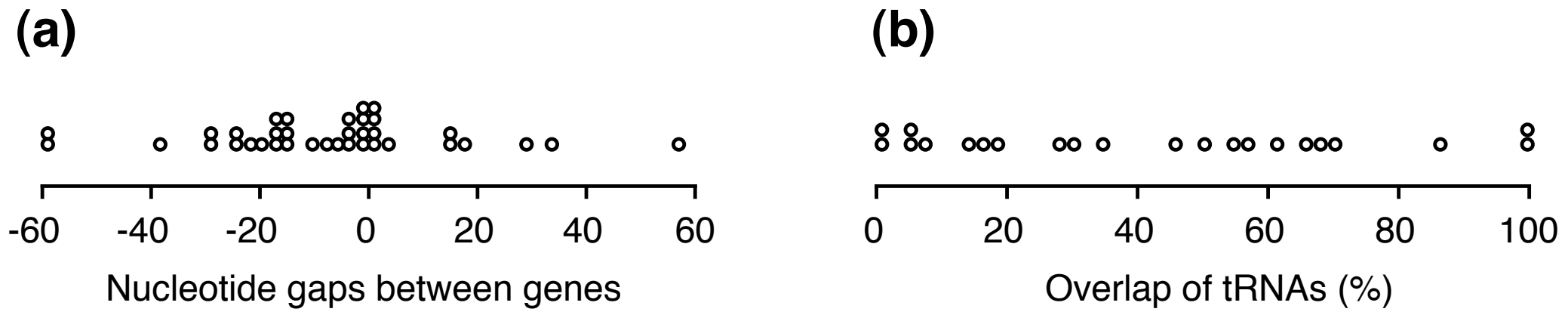

**Fig. 4.** (a) Size of the gaps (number of nucleotides) between adjacent genes, and (b) amount of overlap of each tRNA (%) with another gene, in the *Epiperipatus biolleyi* (Onychophora) mitochondrial genome, based on the new annotations. Each dot represents (a) one boundary between genes and (b) one tRNA. Negative numbers indicate overlap between gene sequences. Not shown is the largest gap between genes (449 nucleotides), which is assumed to be the AT-rich or control region.

There is currently no known biochemical mechanism to account for extensive gene overlap in mitogenomes. However, gene overlap is apparently both common and well-conserved in prokaryotes [38, 39], and so there is presumably no necessary barrier to developing such a concept. Nevertheless, mitochondria are considered to be transcribed as a polycistronic transcript, and there is no published general mechanism by which overlapping genes could then be resolved into whole mRNAs and tRNAs that are gene-specific. Overlap between protein-coding genes is usually ascribed to the use of bicistronic transcripts, as has been shown for Metazoan ATP6-ATP8 and ND4-ND4L [4]. Overlap of tRNA-coding genes is usually ascribed to post-transcriptional editing, such as polyadenylation of the 3' end, as has been shown for several Metazoan tRNAs [49]. However, neither of these possible mechanisms has been experimentally demonstrated except in a small number of specific circumstances, and usually only for very short overlaps (<7 nt). So, they should not be assumed indiscriminately as the explanation for every potential gene overlap that is encountered. Post-transcriptional processing (such as trans-splicing), independent transcriptional promoters and differential transcript cleavage are all possible explanations, and much more experimental evidence is required to understand what is going on in mitogenomes.

### *3.4 Rejecting simultaneous tRNA and rRNA/protein coding*

As emphasized in the previous section, most researchers searched for as-yet unidentified tRNAs only in regions that had not already been assigned as gene-coding [18]. They thus explicitly excluded the possibility that the same location in the mitogenome might code for more than one gene.

However, some authors did consider this possibility, and a few even annotated their mitogenomes with such locations, with a tRNA gene located within the boundaries of either a protein-coding or rRNA-coding gene. Unfortunately, quite a few of these claims were not substantiated by my comparative analyses, in the same manner as discussed in the previous section. My comparative studies showed that most of these claims resulted from lack of consistency between sequence annotations, which could be resolved by adopting a consistent set of annotation decisions across all of the taxonomic groups. A few examples are presented here to indicate the diversity of possible situations; and the relevant alignments are shown in Appendix 3.

One example was discussed in the previous section, that of the tRNA-phe gene of *Aphrocallistes vastus* (Porifera) being annotated as located within the bounds of the NAD5 gene. The re-annotation to remove the gene overlap involved using a truncated TA (instead of the annotated TAA) as the stop codon.

Another example was mentioned in the first section, where the tRNA-asp gene of *Mesobuthus gibbosus* (Arachnida; Scorpiones) was annotated to be located within the bounds of the s-rRNA gene. However, the resulting tRNA sequence matched the model secondary structure no better than would a random sequence.

As a new example, the tRNA-ser(UCN) gene of *Steganacarus magnus* (Arthropoda; Chelicerata) was annotated to possibly be located within the bounds of the l-rRNA gene. The annotated tRNA-ser(UCN) gene sequence aligns unambiguously with the genes of related sequences if 2 nt are added to both the 5' and 3' ends. It then closely matches the model tRNA secondary structure that I have used, but it has a D arm unlike the secondary structure shown by Domes et al. [41] (their Fig. 4). Also, the secondary structure of the final l-rRNA stem shown by Domes et al. [41] (labeled H4 in their Fig. 3a), which includes most of the overlap with the tRNA-ser gene, does not match the rRNA secondary structures shown for the Arthropoda mitochondrial l-rRNA sequences on the CRW web site [21]. If the l-rRNA gene is re-annotated to not include the H4 stem then there would be only 6 nt (out of 61 nt) of overlap with the tRNA-ser(UCN) gene. Alternatively, the l-rRNA gene could be re-annotated to match more closely the CRW H4 stem structures, in which case there would be 19 nt of overlap (plus another ~2 nt if the gene extends beyond the end of the stem). The latter seems to be the most likely option.

The two databased *Varroa destructor* (Arthropoda; Chelicerata) mitogenomes were annotated to have different tRNAs completely overlap with the l-rRNA: in accession AJ493124 the overlap was with the succeeding tRNA-val and in AY163547 it was with the preceding tRNA-pro. In both cases it may be that the alternate boundary of the preceding/succeeding tRNA was used as the boundary of the l-rRNA (i.e. using the 5' boundary instead of the 3' for the preceding gene and the 3' instead of the 5' for the succeeding gene), so that the tRNA was inadvertently included within the annotation ([42] actually state that the tRNA-val "lies next to the rrnL"). However, re-annotating these mitogenomes to match the rRNA secondary structures shown for the Arthropoda mitochondrial l-rRNA sequences on the CRW web site [21], the l-rRNA would overlap with the tRNA-val by 27 nt in order to include the final two l-rRNA 3' stems, which are included in the annotations for all of the related species.

### *3.5 Accepting simultaneous tRNA coding*

Not all mitogenome locations that are claimed to code for more than one gene could be rejected by the comparative analyses. So, I searched the literature for sequences where only one/two tRNA genes were annotated to be missing but none were missing from related species, as this seemed to be the most likely circumstance to locate genuine examples of mitogenomic regions that might have duplicate coding functions. My search was not intended to be exhaustive. I found two distinct types of situation where multiple coding functions were not rejected by my comparative analyses, suggesting that this novel hypothesis is worth considering.

The first situation was that, because of the idiosyncratic structure of Metazoan tRNAs, their genes can be the reverse-complement of each other. That is, they can be coded on opposite strands of the chromosome at the same location. Annotations showing this situation have appeared in the literature [43] but it has apparently not explicitly been commented upon before. The relevant alignments are shown in Appendix 3.

As a first example, the annotated tRNA-cys and tRNA-ala genes of all seven currently sequenced members of the order Mermithida (Nematoda: *Agamermis* sp., *Hexamermis agrotis*, *Romanomermis culicivorax*, *Romanomermis iyengari*, *Romanomermis nielseni*, *Strelkovimermis spiculatus*, *Thaumamermis cosgrovei*) were almost perfectly the reverse complement of each other (Fig. 5). On average, the tRNA secondary structures matched my model as well as did any of the other tRNA genes that I investigated; and these locations were the best-matching ones for these genes within each mitogenome. However, the actual location within each mitogenome was not conserved across any of the seven species (indeed, there was very little conservation of gene positions across these particular genomes), and so the conservation of this arrangement seems to have survived a great deal of genomic rearrangement within the Order. There are no publications associated with any of these genomes, and so this situation has apparently not previously been discussed in print. It might be possible, for example, that incomplete genome amplifications are responsible for these patterns.

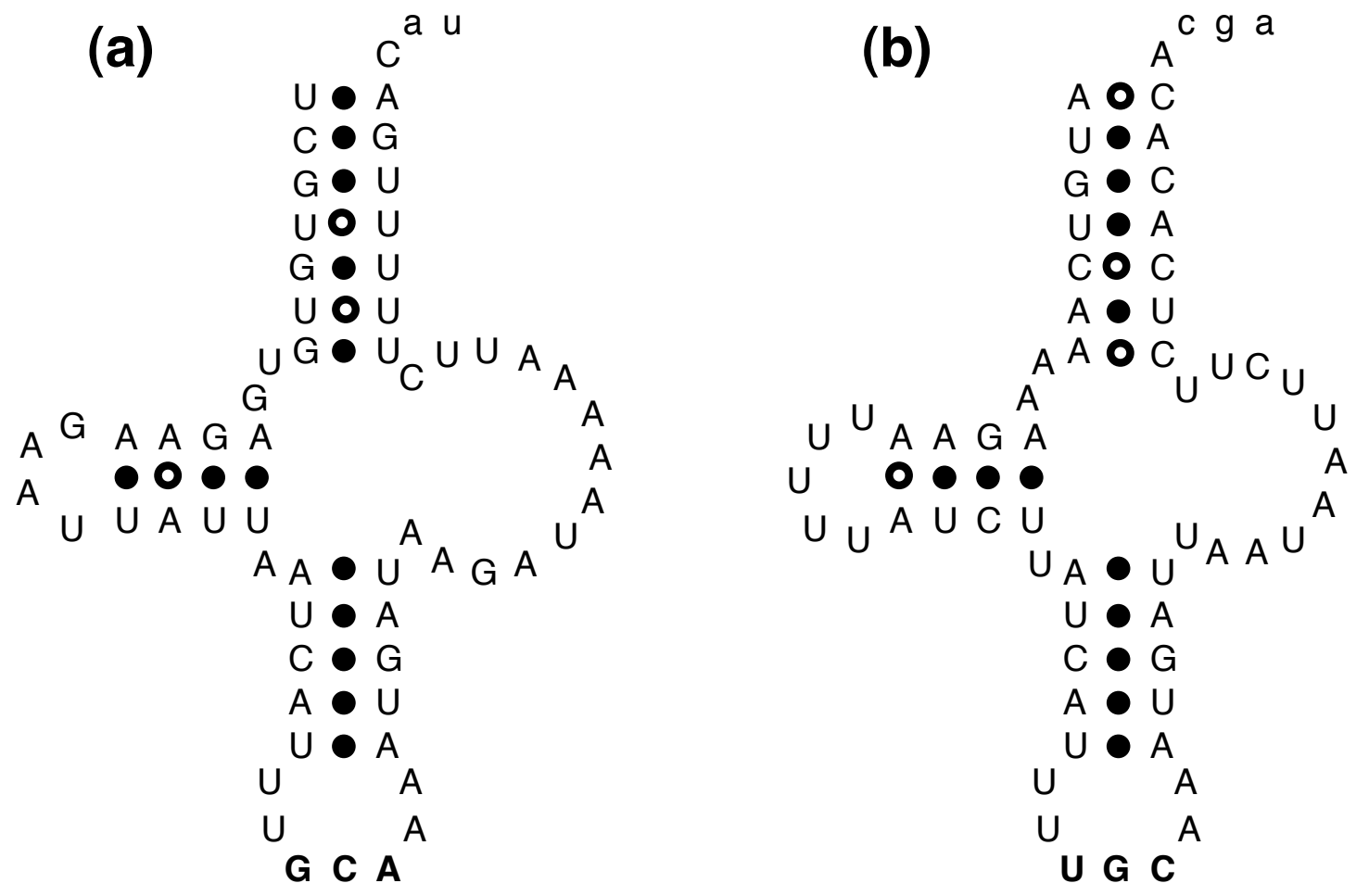

**Fig. 5.** The secondary structures of the tRNA genes of *Thaumamermis cosgrovei* (Nematoda) for (a) Cysteine and (b) Alanine. The filled dots represent canonical base-pairs in the stems and the open dots inferred non-canonical base-pairs; the anticodon is in boldface. The two sequences are the reverse-complement of each other, with the lower-case letters not being involved in the tRNA coding.

Two further examples were identified in the Arthropoda, each involving a previously un-annotated gene. In the *Shinkaia crosnieri* (Arthropoda; Crustacea) mitogenome the tRNA-cys and tRNA-ala genes were the reverse complement of each other (Fig. 6d,e), the same as for the nematodes, while for *Aleurochiton aceris* (Arthropoda; Insecta) it was the tRNA-ile and tRNA-gln genes (Fig. 6f,g). As before, these locations were the best-matching ones for these genes within each mitogenome; however, in both of these examples the related species that have currently been sequenced did not show any evidence of having tRNAs on opposite strands of the same location.

For *S. crosnieri*, the tRNA-ala gene sequence was very similar to that of the related species but the tRNA-cys gene sequence was quite different in the alignment, except in the Anticodon stem. For *A. aceris*, the tRNA-gln gene sequence was the same as that of the related species but the tRNA-ile sequence was quite different in the alignment, including the Anticodon stem. The implied secondary structure of all four tRNAs matched the model structure quite well (Fig. 6). The most likely scenario is thus that the location of one of the paired tRNA genes has been co-opted to simultaneously perform a second role (associated with loss of the original gene), so that one of the paired gene sequences is not homologous with the same gene sequence in the related species.

Another example was annotated (but not discussed) by Thao et al. [43] for the tRNA-trp and tRNA-ser(UCN) genes of *Neomaskellia andropogonis* (Arthropoda; Insecta). Here, the tRNA-trp gene sequence was the same as that of the related species, but the tRNA-ser(UCN) sequence was somewhat different as it lacked several motifs. The implied secondary structure of both tRNAs matched the model structure quite well.

A more complicated example involved two previously un-annotated genes for *N. andropogonis*. The best-matching locations within each mitogenome for the tRNA-cys and tRNA-asn genes were the same location, so that the implied secondary structures were different foldings of exactly the same sequence (Fig. 6 a,b). Furthermore, the best-matching location for the tRNA-ala gene was the reverse-complement of this sequence (Fig. 6c). It thus seems possible that this mitogenomic location may perform three distinct coding functions. I have not identified this situation in any of the other mitogenomes that I investigated, and it remains as a rather intriguing suggestion.

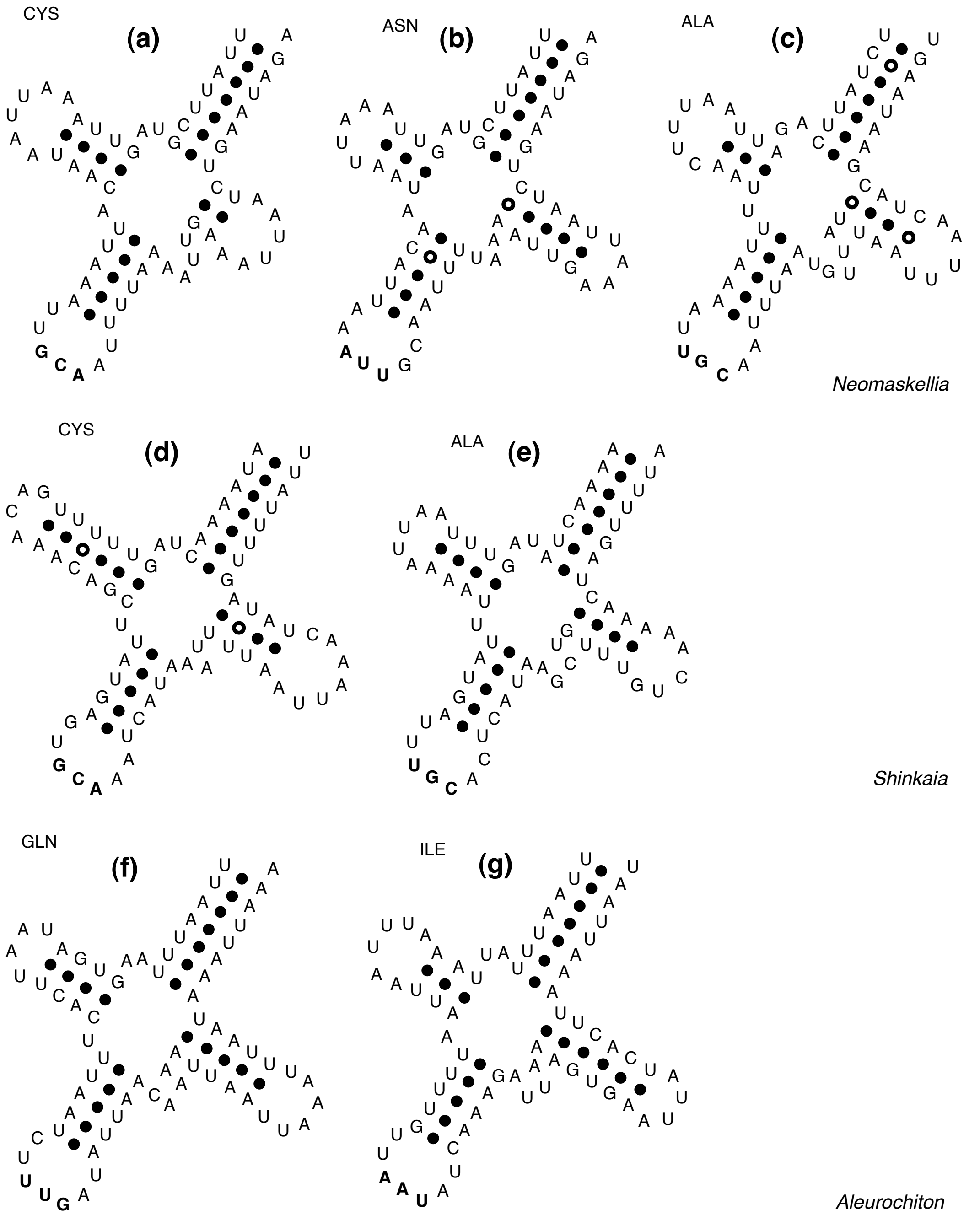

**Fig. 6.** The secondary structures of the tRNA genes of *Neomaskellia andropogonis* (Arthropoda; Insecta) for (a) Cysteine, (b) Asparagine and (c) Alanine; the tRNA genes of *Shinkaia crosnieri* (Arthropoda; Crustacea) for (d) Cysteine and (e) Alanine; and the tRNA genes of *Aleurochiton aceris* (Arthropoda; Insecta) for (f) Glutamine and (g) Isoleucine. The filled dots represent canonical base-pairs in the stems and the open dots inferred non-canonical base-pairs; the anticodon is in boldface.

As before, the implied secondary structure of all three tRNAs matched the model structure quite well (Fig. 6). Furthermore, the tRNA-cys gene sequence was very similar to that of the related species, but both the tRNA-asn and (particularly) the tRNA-ala gene sequences were quite different in their alignments. Once again, this suggests that the location of one of the tRNA genes has been co-opted to simultaneously perform the other two roles.

If the examples of multiple coding hypothesized here are correct, then the most obvious explanation for the situation is selective pressure for smaller mitogenomes, just as extensive gene overlap is considered to result from selection for smaller microbial genomes [44]. The situation is apparently a derived but polyphyletic characteristic in Metazoa, since I encountered it only sporadically among the taxonomic groups that I examined. However, I recorded it twice in *N. andropogonis*, which suggests that there may be some conditions that predisposes certain genomes.

The alternative explanation, used by most of the original authors, was that some of the tRNA genes are missing from certain species but not from their relatives; that is, that loss of a small number of genes is a recurrent phenomenon within the Metazoa. These genomes would then presumably rely on a nuclear copy of the relevant tRNA genes, with the tRNA being imported into the mitochondrion [45]. Testing of the these two alternative hypotheses will require analysis of the processed tRNA transcripts, as would the other possible explanation of tRNA gene recruitment [46].

Perhaps the biggest practical problem associated with this situation for Metazoan mitogenomes is that the ARWEN and DOGMA programs usually identified only one of the paired tRNAs. For example, for *S. crosnieri* ARWEN identified tRNA-ala and DOGMA identified tRNA-cys, for *A. aceris* ARWEN identified tRNA-ile and DOGMA identified tRNA-gln, and for *N. andropogonis* ARWEN identified tRNA-cys and DOGMA tRNA-asn and tRNA-ala. This re-emphasizes the need to use both of these programs when trying to locate the full complement of tRNAs in mitochondrial genomes.

### *3.6 Accepting simultaneous tRNA and protein coding*

The second situation where multiple coding functions were not rejected by my comparative analyses is perhaps more remarkable, and involved tRNA-coding genes that were inferred to occur within the boundaries of a protein-coding gene, usually on the same strand. This situation has been reported in passing before [47] but all of these claims that I have encountered in the literature could be resolved by re-annotation, mainly because they were near the ends of the genes. The examples discussed in this section all occur in the middle of the protein-coding region, so that this simple form of re-annotation is not feasible.

For the comparative analyses of the structure model of the tRNA, putative tRNAs were required to have the Anticodon stem with canonical pairs only and to have (almost) exactly the same nucleotide motifs as did the related sequences. The D and/or T arms often had poorly paired stems, and were sometimes of different length to those of related species, or did not have motifs that were conserved in the other sequences. The Acceptor stem usually fitted the model secondary structure well (e.g. only one non-canonical pairing). The relevant alignments are shown in Appendix 3.

The first example concerned *Xiphinema americanum* (Nematoda), in which the tRNA-asn gene was located within the coding region of the COX2 gene (Fig. 7 and Appendix 3). The tRNA alignment was no worse than any of the others in my study, as it contained phylogenetic motifs as well as matching the model tRNA structure. However, the COX2 alignment had several unique modifications that affected the inferred amino acid alignment (Fig. 7). The location of the tRNA Anticodon loop, in particular, was at odds with a region of the COX2 that was highly conserved across all of the nematodes (Fig. 7). It thus seems that the amino acid coding has been compromised in order to accommodate the simultaneous coding of the tRNA. This presumably results from competing selection pressures, so that it is impossible to maintain both coding sequences perfectly. It also imples that the tRNA gene is not homologous with the same gene sequence in the related species.

```
                                                   tRNA-Asn
                   Structure:   ((((((--((((((----------))))))-(((((--***--)))))---------)))))))-
Strelkovimermis spiculatus      TTTAAGCAT------TAT------G------ATAAAACTGTTAATTTTATTGAAAAAAGCTTAAGA
Agamermis sp. BH-2006           ....G.TTA------A--------A------TG....T.........C.AC....TTA.C..GAT
Hexamermis agrotis              A.....TG.-------..A------T------G.G...T.........C..GT..TTT-A....GAG
Thaumamermis cosgrovei          CA....TTC------A.A------C------.....T.....C.......AG..T.-.......T
Romanomermis nielseni           .....A..-------CTATA----A------.....T............C....--.T....A.
Romanomermis culicivorax        A....A.G.------ACAA-----T------.....T............T....--.T....A.
Romanomermis iyengari           .....A..A------AGAA-----T------.....T............A....--.T....A.
Trichinella spiralis            .AG.GT..AGCAT--ATCAGAT--.--ATGC...........T.....AGA.T.C.-AAC.CTAG
Xiphinema americanum            .GGTTCTTCAAGG--GTG------T--CTATTGGCTCT.....TGAG.TG..CGG..-CTGGCCAC
---------------------------------------------------------------------------------------------------
Ascaris suum                    .AAG..TTAGTTT--A..ATT---A--GAATTGTTG......G..CAATGG.TG.T--T..CTTAG
Anisakis simplex                .AAGGTTTAGTTT--A..TTT---A--GAATTGTCG......G..CGG.AG.TTTT--TACCTTAG
Toxocara canis                  .AG...TTAGTTT--A..AA----A--AAATG.TTG.........CAAGAGAC.T.--T...CTAG
Toxocara malaysiensis           .AG...TTAGTTT--A..AAG---A--AAATG.TTG.........CAA.AGAC.TG--T...CTAG
Toxocara cati                   .AG...TTAGTTT--A..AAG---A--AAATG.TTG.........CAA.AGAC.T.--T...CTAG
Necator americanus              .AAG..TTAGTTT--A..A-----A--GAATTGTTG.........CAA.AG.TTT---A..CTTA.
Ancylostoma duodenale           .AAG..TTAGTTT--A.GTGTGTTA--GAATTGTTG.........CAA.AG.TTT---A..CTTA.
Cooperia oncophora              .AAGG.TTAGTTT--A.GT-----A--AAATTGTTG.T....G..CAA.AGATT.---T..CTTA.
Haemonchus contortus            .AAG.ATTAGTTT--A..TTA---A--GAATTGTTG.........CAA.AG.TG.---TT.CTTA.
Caenorhabditis elegans          .AAG..TTAGTTT--A..AT----A--GAATTGTTG.........CAA.AG.TGT.--C..CTTA.
Caenorhabditis briggsae         .AAG..TTAGTTT--A..TAA---A--GAATTGTTG.........CAA.AG.T.T.--C..CTTA.
Heterorhabditis bacteriophora   .AAG..TTAGTTT--A..T-----A--GAATTGTTG......G..CAA.AG.TT.---A..CTTA.
Steinernema carpocapsae         .AAG..TTAGTTT--A..TT----A--GAAT.ATTG.........CAATAG.TTTT--T..CTTA.
Strongyloides stercoralis       .GAGGATTATTTT--A..TT----A--GAAT..TT.G.......C.AA.AG.TT----ATC.TCAT
Enterobius vermicularis         .AGGCTGTAGTTT--G..AGA---A--AAATGGTGG.T....GC.CCG.CGA..---TGGCCTAG
Globodera pallida               AAAG.TTTATTTA--A.AA-----A--TAAT..TTT.T.......AAA.AGTTTTT--T...TTTT
Globodera rostochiensis         AAAG.TTTATTTA--A.AA-----A--TAAT..TTT.T.......AAA.AGTTTTT--T...TTTT
Onchocerca volvulus             GAATTTTTAGTAT--A..TTTT--.--GTACGGTT..T....G...AGCAG.TTTT--.AAGTTCT
Dirofilaria immitis             GAATTTTGAGTATG-GT.T-----T-TGTAT.ACT..T....G...AGTAG.GT.---AAGGTTCT
Brugia malayi                   GAATTTT.AGTATAA.T.TT----TTTGTAT.ACT..T....G...AGTAG.TT.---AAGATTCT
```

```
                             Cox2                                                    Cox2
312312312312312312312312312312312312312312312312312312312312312123123
AAGAAATTTACCAAGACTTCATATTGGTCAGTGTAGAGAATTGTGCGGAAGCTACCAC          GRLNYYVLSSNLPSLHIGQCSELCGSYHSWMPI
..TTTTGAGTATTG.TAA.T....A..A..A...TC......A..T....AT.....T          ....MLSTKILSIGNYM........N.......
..TTTC...CATTG.CAAA........A..A...TCT...C.A..T..C.A...T..T          ....LLSSQISFIGK..........N.......
..AC...CAG..T..CAAATT.........A.....G...C.......T.A..TT..T          ....FINILN.Q..KF.........NF......
.TTT.GAA.T...G..AAGTT...A..A..A..C.....GC.T..T..G..G.....T          ....F.MMMFSI.GKFM................
GTTT.TGA.....G.TAAATT......A.....C.....GC.A..T..G.CA..T..T          ....F.MMMFMM.GKF.........T.......
.TTT.G.A.T...G..AAATT.T....G..A..C......C.A..T..TGTT..T..T          ....F.M.MFSI.GKFF........V.......
.TCC.CAC.T...G..G.CAT.TA...A..A...TC....C.T..T..CGTAA.....          ..I.STTTE.T..GVIY........VN..F...
TG.T---.CTT....GG.GTC......CTCT...TAT..G..........CTGG....          ..I.STIIDG-SS.VS..S.Y....TG..A..V
--------------------------------------------------------------------------------------------------
TTTTCC.G.GGTTG.TG.ATT.TA......A...TCT...A.T..T..TGCTA.T..T          .I.STLSY.FPVVGVFY.....I..AN..F..V
TTTTCC.ACTGTGG.TG.GTT.TA......A...TC...GA.T..T..GGCTA....T          .I.STVSY.FPTVGVFY.....I..AN..F...
TTTTCC.G..GT.G.TG..TT.TA...G..A...TCT..GA.C..T..TGCTA....T          .I.TTLSY.FPVVGVFY.....I..AN..F...
TTTTCC.G.GGTGG.TG..TT.TA..........TCT..GA.T..T..TGCTA....T          .I.TTLSY.FPVVGVFY.....I..AN..F...
TTTTCC.G..GT.G.TG..TT.TA......A...TCT...A.C..T..TGCTA.....          .I.TTLSY.FPVVGVFY.....I..AN..F...
TTTTCC.G..GTGG.GG..TT.TA..........TC...GA.T..T..TGCTA.T..T          .I.STLCY.FPVVGVFY.....I..AN..F...
.TTTCC.G..GT.G.GG..TT.TA......A...TCT..GA.T..T..GGC.A.T..T          .I.STLCY.FPVVGVFY.....I..AN..F...
.TTTCC.A.GATTG.TG.ATT.TA......A...TCG...A.T..T..GGCTA.T..T          .I.STLCY.FPMIGVFY.....I..AN..F...
.TTTCCGC.GATTG.GG.GTT.TA......A...TCG...A.T..T...GCTA.T..T          .I.STLCY.FP.IGVFY.....I..AN..F...
GTTTCC.A..GTGG..G..TT.TA......A...TC...GA.T..T...GCAA.T..T          .I.STFSY.FPMVGVFY.....I..AN..F...
TTTTCC.A.GGTGG.GG.ATTCTA...A..A...TC...GA.T..T...GCTA.T...          .I.STFSY.FPMVGVFY.....I..AN..F...
.TTCCC.ACTGTTG..G..TT.TA..........TCT..GA.T..T..TGCTA.T..T          .I.STLCY.FPTVGVFY.....I..AN..F...
TTTTCC.A.GATTG.TG..TT.TAC..C......TCT...A.T.....TGC.A....T          .I.STISY.FPMIGVFY.....I..AN..F...
TTTCCC.A.GG.TG.TG..TT.TA.......A..TCT...A.T..T..TGCTA.T..T          .I.STVTC.FPMAGVFY..S..I..AN..F...
GTTTG....TGTTG.TG..T..TA....AT....TC...GA.T.....GGCTGGT..T          .IN.SHL.WFDNVGVYY.M...I..AG..Y..V
TTTTG....GGTTG.GG..TT.T....G..A...TCG..GA.T.....GGCAA.T..T          .L..ILNFNFD.VGVFF.....I..AN..F..V
TTTTG....GGTGG.GG..TT.T....G..A...TCG..GA.T.....GGCAA.T..T          .L..ILNFNFD.VGVFF.....I..AN..F..V
TTTTTC..GTT.TG.TT.GTT.T....G......TCT...A.T..T..TGCTA.T..T          .L.TKVTC.FSCSG.FF.....I..AN..F...
TTTTTC..GTT.TG.TT.ATT.TA..........TCT..GA.T..T..GGCTA.T..T          .L.TKITCNFSCSG.FY.....I..AN..F...
TTTTTC..GTTGTG.TT.GTT.TA...C..A...TCT..GA.T..T..TGC.A.T...          .L.TKITFNFSCCG.FY.....I..AN..F...
```

**Fig. 7.** Alignments of the tRNA-asn and COX2 genes of *Xiphinema americanum* (Nematoda) with the other nematode mitogenomes. The translated amino acids are shown along with the two nucleotide sequences. The sequences are arranged so that taxonomically close species are near each other; dashed line = Dorylaimia (above) and Chromadoria (below). Underline = *X. americanum* tRNA-asn sequence; double underline = anticodon loop; dots = match to the first sequence. Structure: () = stem pairs; * = anticodon; numbers = codon positions.

A somewhat different situation concerned *Shinkaia crosnieri* (Arthropoda; Crustacea), in which the tRNA-trp gene overlapped part of the coding regions of both the COX1 and CYTB genes on the opposite strand (Fig. 8 and Appendix 3). The tRNA-trp sequence aligned quite well with the related sequences, but the alignment based on the structure model either (1) was shifted 1 nt from the phylogenetic alignment (my preferred reconstruction) or (2) the D stem and T stem each had 1 less pairing (an alternate reconstruction). There was also a much longer T-loop, which coincided with the boundary of the CYTB and COX1 genes. The COX1 gene is usually the most highly conserved one in mitochondria, and yet the 3' end of the *S. crosnieri* gene is completely different to those of all of the related species (Fig. 8). The hypothesized dual coding with the tRNA-trp gene would provide a very simple explanation for this situation, just as it does for the *X. americanum* tRNA-asn and COX2 genes. Thus, in these two cases a single hypothesis explains two oddities: the apparent absence of a tRNA-coding gene, and the apparent non-conservation of a protein-coding gene.

A second example of complete overlap concerned *Epiperipatus biolleyi* (Onychophora), in which the tRNA-phe gene was located within the coding region of the COX3 gene (Appendix 3). The amino-acid alignment was very well conserved, but at the nucleotide level the other COX3 sequences would have several mis-pairings in the anticodon stem, if their transcribed product was to be folded into a tRNA. However, the tRNA gene sequence was quite different to the other sequences in the alignment, other than in the Anticodon loop. It thus seems that it in this case it was the tRNA coding that had been compromised.

A third example of complete overlap also concerned *E. biolleyi*, in which the tRNA-leu(CUN) gene was located within the coding region of the ATP6 gene (Appendix 3). At the nucleotide level the protein alignment was quite well conserved, although not so well at the amino acid level. Once again, the tRNA gene sequence was quite different to the other sequences in the alignment, other than in the Anticodon loop. Moreover, in this case the T stem was apparently missing, which seems to be a much greater structural compromise, since this requires a different translation mechanism [48]. This implies stronger selection pressure. The length of the T stem in the other aligned sequences was variable (2–5 pairs), as it often is in the Metazoa.

The fourth example concerned *Aleurodicus dugesii* (Arthropoda; Insecta), in which the tRNA-gln gene was located within the coding region of the NAD5 gene (Appendix 3). However, in this case the two genes were on opposite strands, and the coding of both genes seems to have been compromised. The region of the NAD5 sequence was a very variable section as far as the amino acids were concerned. However, the nucleotide sequence differed in three highly conserved columns, one in the Anticodon loop and one in the D stem (the other was unpaired). The t-RNA sequence matched the related species only in the Anticodon stem, and the structure lacked the T arm entirely.

If the dual coding that I have hypothesized here is correct, then the most obvious explanation for the situation is selective pressure for smaller mitogenomes, in the same way that loss of the D or T arm alone is considered to have resulted from selection for smaller mitogenomes [44]. As for the situation with multiple tRNA coding (discussed in the previous section), this is apparently a derived but polyphyletic characteristic in Metazoa. Similarly, one of the two genes in the multiple-coding region (presumably the tRNA) is not homologous with those of closely related species.

```
                                                                    tRNA-Trp
                  Structure:   ((((((--((((-------))))-(((((--***--)))))----(((((----------------)))))))))))-
Consensus                      AAGATTTTAAGTTATTTTAAAACTAAAAGCCTTCAAAGCTTTCAATAAGAGTAAAAAT--------TCTCTTAAGTCTTA
Gonodactylus chiragra          ....CC.......A.A--.........A......C........T....C..--------------......G......
Lysiosquillina maculata        .....C.......AC.--......G..........C........T....TC...---------C......G......
Harpiosquilla harpax           .....C.......AACG-......G...A......C...C.T...T.....T..--------------......G......
Squilla empusa                 .....C.......AAC--......G...A......C...C.....T.....C..---------C...C..G......
Squilla mantis                 .....C.......AA.--......G...A......C...C.....T.....TT..-------------......G......
Ligia oceanica                 T.A.C........AC.A.......TT.T........A.AAA.T.--..-.C.TCT.--------.-.--.....T.A.
Fenneropenaeus chinensis       ..AGC.........C.--.....T.T............A....A.G........-----------....C....CT...
Litopenaeus vannamei           ..AGC.........C.--.......T..........T.A....A..........---------------.......CT...
Marsupenaeus japonicus         ..AGC...........--......GC............GC...A.G.......T----------------.......CT...
Penaeus monodon                ..AGC.........CA--.....T.T............G....G.G.....G..----------....C....CT...
Panulirus japonicus            ................-.....GTTGC.........GCGAA..A..A..ACGT-----------G..T....A.....
Cherax destructor              .............A..-.......C.T...........G.......A...TG..----------C..T....A....G
Callinectes sapidus            .G..............--........G..........C.AA..A..A..ATC.-----------....T....A..C..
Portunus trituberculatus       .G............A.--........G.........TC.AA..A..A..A.TTC----------....T....A..C..
Geothelphusa dehaani           .............AA.--......G...............AA...--A..-.TTTT----------.-.T--..A.....
Pseudocarcinus gigas           .GA.............--......................AA..A..A......C----------A..T....A.TC..
Eriocheir sinensis             .GA.............--..........................T.A..A....TT..---------C..T....A.TC..
Halocaridina rubra             ...G.........AAA--......GT.A........T..CT....G...A..GC----------A.......AC....
Macrobrachium rosenbergii      ...G.........AC.--......GT.A........T.CC...C.......T..--------------.......AC....
Pagurus longicarpus            G.............A.--.......T.............A.A..A.TA......C-----------....T....A.....
Shinkaia crosnieri             T.A.A.A...AG.-.CC---.TT....T.TT......A.A..TGT.TTA.A....CTAAATAATCA.T.TAA..T.T.AT
```

```
      Cytb                                         Cox1                               Cytb          Cox1
23123123123123123    1231231231231231231231231231231231231231231231231231231                             
ATAAAATACTAAAATAA    ACGCAACGATGATTATTTTCTACAAATCATAAAGAYATTGGAACATTATATTTTA     WDKMLK*  TQRWLFSTNHKDIGTLYFIFG
..G....C..T.G....    ....G...........................C.....C..G..........     ..EI.S.  .R.................L.
.CG....T..T.GC...    ....G..................C..C........T.....T.............     ..EI.S.  .R.................L.
.CG.G..T..T.G....    ....G...........C.......C........T..C.....TC.........     ..EI.S.  .R.................L.
.CG....C..T.G....    ....G...........................T.......TC.......C.     ..EI.S.  .R.................L.
.CG....T..T.G....    ....G...........................T.......T........C.     ..EI.S.  .R.................L.
.C......A.T---...    ....G...T...C.T.A.....C..C.....G..T.........C........G     ....I-.  .R...Y............V..
...........G.T...    ................................C..C......C....C....     .....D.  .....................
...........G.T..-    ..........................C..C.....C..............C....     .....D.  .....................
...........G.T..-    ................C.......C.......C.........GC........     .....D.  .....................
....G......G.T..G    ................................C..C.....TC.........     .....D.  .....................
..T.TC.C...G.C.--    .............T..C.....T..........C.....T..C.....C..C.     ..YL.D.  ....F................
.CTCCC.T..T....--    .............T.....A.............T........C.......CG     T.SL...  ....F.............V..
.......GT......--    .T...........C...................C.....T..........C.     .......  M...F................
...............--    .T...........C...................T.....T.............     .......  M...F................
........T......--    .T...........T.....C.....C........T..C..C..T..........     .......  M...F...............L.
....G...T.......-    .T...........C...................C.....T.............     .......  M...F................
....C.....T....--    .T..........C.T..................T....................     ..N....  M....................
.....C.....G.T...    C.......T...C.C..C..C..C..C........T........TC.T..C..C.     ...L.D.  P..................L.
.C...C.CA..G.C.--    ........G..GC.T..C..A.....C..........C.........C.........     ...LMD.  .....................
........AA.CTT...    ...A.........T..C.....T..C..C.....C.........C.........     ....KL.  .K..F................
.......TT......G.TTATTTAGTTTAT.T.AA.CAAATGTTTTGAA.AC.TTTA.A......T..........     ..I...--LVYLKQMFWKHLN.......
```

**Fig. 8.** Alignments of the tRNA-trp gene and parts of the COX1 and CYTB genes of *Shinkaia crosnieri* (Arthropoda; Crustacea) with other Crustacea mitogenomes. The translated amino acids are shown along with the nucleotide sequences. The sequences are arranged so that taxonomically close species are near each other; symbols are as in Fig. 7.

It is unclear what biochemical mechanism would allow alternate transcription leading to the tRNA rather than to the mRNA. However, throughout the Metazoa some genes have been observed to overlap by a variable (sometimes large) number of bases (see above), so there must be some general mechanism for recognizing alternate transcription start/stop sites. Once again, experimental testing of my hypothesis will require analysis of the processed tRNA transcripts.

It is important to note that the conclusions drawn here rely very much on the quality of the comparative analysis, in particular the availability of sequences from closely related species. Otherwise, it is possible to get "carried away" and find potential dual-coding regions in all sorts of improbable places. One example that I encountered concerns *Buthus occitanus* (Arthropoda; Chelicerata), which is currently annotated to lack the tRNA-asp gene. There is a location within the coding region of the NAD1 gene that, using an alignment based on the sequences of Masta and Boore [17], seems to be a candidate for dual coding. However, this situation is not repeated in the sequences of any of the other members of the Buthidae (*Centruroides limpidus*, *Mesobuthus gibbosus*, *Mesobuthus martensii*), which are also annotated to lack the tRNA-asp gene. In this case, it is better to hypothesize the true lack of a mitochondrial tRNA-asp gene in this family, as this is a more consistent explanation.

As an alternative example, *Shinkaia crosnieri* is currently annotated to lack the tRNA-ser(UCN) gene. There is a location within the coding region of the NAD6 gene that could be a candidate for dual coding. However, not only is the tRNA-ser gene different to those of related species but the location in the NAD6 is the least-conserved part of that gene, so that the alignment is very uncertain. Under these circumstances, it is rather difficult to justify a claim for dual coding, due to rather poor bioinformatic evidence.

Finally, tRNA-like structures have previously been reported totally within protein-coding genes even when there is a standard tRNA gene in the mitogenome, for example in the snail *Albinaria coerulea* [49]. Here, the authors considered the tRNA-like structure to be non-functional.

## *3.7 tRNAs without D or T arms*

Okimoto and Wolstenholme [50] raised the possibility that in some organisms mitochondrial tRNAs might lack both the D arm and the T arm, given that they can lack one or the other in many members of the Metazoa (and that the stems are variable in size even when they do exist). It is difficult to address this possibility by examining gene sequences, because searching for locations where the genes would imply the required tRNA structure (i.e. a pattern-matching approach) is likely to be futile.

To illustrate this point, the mitogenome of *Ligia oceanica* (Arthropoda; Crustacea) contains a location (at one end of what is annotated as the control region) which can be folded into three distinct structures (Fig. 3): a D-armless tRNA-arg, a D-armless and T-armless tRNA-arg, and a single stem-loop. The latter is its current annotation [51], although there is no other suitable location in the genome for the tRNA-arg gene. This apparently contradictory situation is not unexpected, given the strong nucleotide biases present in mitochondrial genomes, and there will thus be many regions that can be folded into stem-loops. The simple structure of a D- and T-armless tRNA (basically a stem-loop with 7-nt loop and a large internal bulge; Fig. 3b) is particularly likely to be found in several locations in any given mitogenome.

Nevertheless, I encountered one potential suite of examples where the hypothesis of D- and T-

armless tRNAs was not rejected by my comparative analysis. That is, there was a set of related species all sharing the same pattern, thus providing the consistent comparative analyses that I am requiring. This situation concerned one particular group of nematodes, the Dorylaimia. (The Enoplea group in the NCBI Taxonomy is treated here as consisting of the Dorylaimia plus Enoplia; there are currently no Enoplia mitogenomes in GenBank.)

As my initial example, *Xiphinema americanum* is currently not annotated as having a gene for tRNA-cys [52]. Moreover, a manual search showed that there is only one region (currently unannotated) that can be folded into the Anticodon stem-loop of a tRNA-cys. This location can also be folded into an Acceptor stem (Fig. 9), but there are no available nucleotide pairs that could form either a D stem or a T stem. The TV-replacement loop of this structure is longer than the D-replacement loop, which would allow the secondary structure to fold into the tertiary "L" structure used by tRNAs. The alignment of this region with the 28 annotated tRNA-cys gene sequences from the Nematoda does not preclude the possibility that it is a tRNA gene, for a reason that needs to be discussed in some detail.

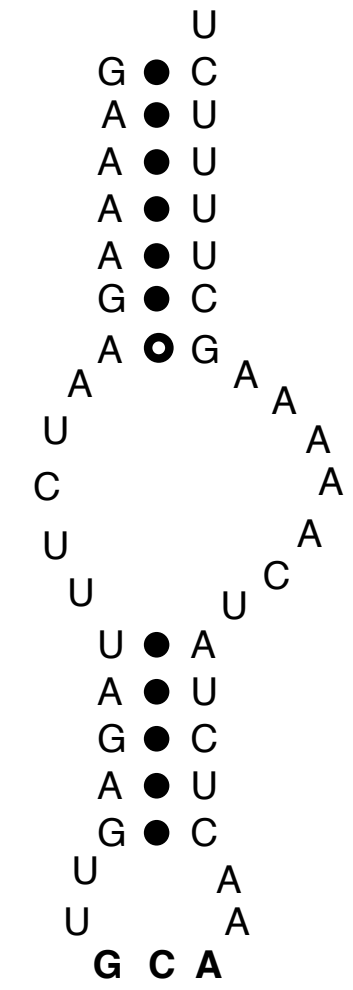

**Fig. 9.** Sequence and implied structure of the hypothesized tRNA-cys gene of *Xiphinema americanum* (Nematoda). The filled dots represent canonical base-pairs in the stems and the open dots inferred non-canonical base-pairs; the anticodon is in boldface.

Within the Nematoda, the alignment of the tRNA genes for Asp, Gln, Glu, Gly, Leu(CUN), Leu(UUR), Met, Pro and Trp was unproblematic in my comparative analyses. That is, ARWEN, DOGMA and the GenBank annotations all agreed on the mitogenome locations, and there were clear phylogenetic motifs in the alignments. The tRNA genes for Lys, Ser(AGN) and Ser(UCN) required some re-annotation (i.e. realignment, or choice of a different part of the genome as the coding sequence; see Appendix 2) to get clear phylogenetic motifs throughout the Nematoda, but this was still a relatively straightforward analysis.

For the remaining 10 tRNA genes, there were problems within the Dorylaimia (but not in the other nematode group, the Chromadoria), notably for the seven species of the Mermithida. In particular, the D arm as annotated varied in length in all cases, and in most cases not all of the stem-pairs were canonical. Similarly, not all of the pairs in the Acceptor stem were canonical for any of these tRNAs. Indeed, there was more sequence variation among the three *Romanomermis* species than among the whole of the Chromadoria.

Furthermore, ARWEN, DOGMA and the GenBank annotations frequently disagreed on the location of each gene (if they found it at all), what particular anticodon it had, or which (if either) arm was replaced by a loop; and the annotations (which usually had consistent Anticodon stems and loops, except for tRNA-phe) did not have consistent phylogenetic motifs. The problems with the secondary structure of the gene locations suggested by ARWEN and DOGMA were that many of them either: (i) had two arms; (ii) were missing the "wrong" arm compared to the other nematode sequences; (iii) had high scores from lots of stem-pairs in the D or T arms but not in the Acceptor or Anticodon stems; (iv) had an unusual anticodon; or (v) were in an rRNA-coding gene, a protein-coding gene, or had multiple tRNAs in the same piece of sequence. The latter possibilities have been discussed in detail in a previous section.

For the Mermithida species, it was possible to resolve or partly resolve the apparent problems by excluding the requirement for a D arm in the tRNA structure model. Since most nematode tRNAs do not have a T arm, the model reduces to that of a D- and T-armless tRNA, as discussed for the Cys gene of *X. americanum*. The full alignments for the tRNA-cys gene sequences are shown in Appendix 5, as an illustrative example, and the improvements in stem-pairing are listed in Table 1; abbreviated data for the other genes are shown in Appendices 5 and 6.

Employing this approach to resolving the alignment inconsistencies worked very well for the tRNA genes of Ala, Asn, Cys and Ile. This greatly improved the phylogenetic motifs, and there were characteristic motifs for taxonomic subgroups. The pairing in the Acceptor stem was improved, and the unalignable loops were also shortened. There were compensatory base changes among the motifs, and a consistent structure between stems. Furthermore, the re-annotation reduced the overlap between genes by an average of 3.1 nt (SD = 4.4).

Furthermore, this approach somewhat improved the motifs and Acceptor pairing of the tRNA genes for Arg, His, Phe and Val. For tRNA-his, the gene sequences of *Agamermis* sp. and *Thaumamermis cosgrovei* were different to all of the others, but these were the best locations for these genes in their mitogenomes. For the tRNA-phe gene, the *T. cosgrovei* sequence was also different to all others, but once again this was the best location for this gene. For tRNA-his, I could realign the sequences to find improved phylogenetic motifs in 5 of the 7 sequences, but not in either *Agamermis* sp. or *Thaumamermis cosgrovei*. For tRNA-val, this approach also somewhat improved the phylogenetic motifs, but the stem-pairing was not improved because it was already good in the orignally annotated alignment.

Unfortunately, it proved to be very hard to find consistent phylogenetic motifs in the tRNA genes for Thr and Tyr, using any approach at all. However, at least for tRNA-thr the annotated motifs were poor only for the three *Romanomermis* species. These two genes are thus intriguing (and frustrating), and appear to require some alternative explanation for their lack of phylogenetic motifs and canonical Acceptor-stem pairing. Perhaps they are pseudogenes, although this seems to be unlikely for the reasons given by Masta and Boore [17, 18] (i.e. canonical pairing in most of the stems, and conservation of motifs in the anticodon stem and loop, notably the anticodon).

Of all the novel hypotheses proposed in this paper, this is the most problematic one, because the lack of both D and T arms apparently conflicts with what is known about the functioning of tRNAs. The roles of the nucleotides that connect the Acceptor and Anticodon stems are to stabilize the tertiary structure and to promote interactions with the proteins involved in translation [48]. When genes were first observed for tRNAs without a T stem they were called "bizarre" [53], since interactions between the D- and T-arm loops were considered necessary for structural stability of the tRNA, but these genes were soon shown to be functional [50]. However, it took quite some time before the required functional mechanism of the tRNAs themselves was elucidated. This requires two thermo-unstable elongation factor (EF-Tu) molecules, one for tRNAs without a D arm and one for those without a T arm [54; reviewed by 48]. A tRNA without either a D or a T arm would thus presumably require a third type of EF-Tu; and it is not at all obvious how the two replacement loops would stabilize the tertiary structure of the tRNA.

Table 1. Number of canonical stem-pairs for the implied tRNA structure of the tRNA-cys gene, as annotated in the Nematoda mitogenomes, plus the new Acceptor annotation without a D-arm.

| | Acceptor | D-arm | Anticodon | T-arm | New |
|---|---|---|---|---|---|
| Max no. stem pairs | 7 | 4 | 5 | 0 | |
| CHROMADORIA | | | | | |
| Ascaridida | | | | | |
| *Anisakis simplex* | 6 | 4 | 5 | 0 | |
| *Ascaris suum* | 6 | 4 | 5 | 0 | |
| *Toxocara canis* | 6 | 4 | 5 | 0 | |
| *Toxocara cati* | 6 | 4 | 5 | 0 | |
| *Toxocara malayensis* | 6 | 4 | 5 | 0 | |
| Spirurida | | | | | |
| *Brugia malayi* | 6 | 4 | 5 | 0 | |
| *Dirofilaria immitis* | 6 | 4 | 5 | 0 | |
| *Onchocerca volvulus* | 6 | 4 | 3 | 0 | |
| Strongylida | | | | | |
| *Ancylostoma duodenale* | 7 | 4 | 5 | 0 | |
| *Cooperia oncophora* | 6 | 3 | 5 | 0 | |
| *Haemonchus contortus* | 7 | 3 | 5 | 0 | |
| *Necator americanus* | 7 | 4 | 5 | 0 | |
| Rhabditida | | | | | |
| *Caenorhabditis briggsae* | 7 | 4 | 5 | 0 | |
| *Caenorhabditis elegans* | 6 | 4 | 5 | 0 | |
| *Heterorhabditis bacteriophora* | 6 | 4 | 5 | 0 | |
| *Steinernema carpocapsae* | 7 | 4 | 5 | 0 | |
| *Strongyloides stercoralis* | 7 | 4 | 5 | 0 | |
| Oxyurida | | | | | |
| *Enterobius vermicularis* | 7 | 4 | 5 | 0 | |
| Tylenchida | | | | | |
| *Globodera pallida* | 7 | 3 | 5 | 0 | |
| *Globodera rostochiensis* | 7 | 3 | 5 | 0 | |
| DORYLAIMIA | | | | | |
| Trichinellida | | | | | |
| *Trichinella spiralis* | 7 | 3 | 5 | 0 | |
| Dorylaimida | | | | | |
| *Xiphinema americanum* | 6 | 0 | 5 | 0 | |
| Mermithida | | | | | |
| *Agamermis* sp. | 6 | 3 | 5 | 0 | 6 |
| *Hexamermis agrotis* | 5 | 3 | 5 | 0 | 7 |
| *Romanomermis culicivorax* | 4 | 3 | 5 | 0 | 6 |
| *Romanomermis iyengari* | 6 | 3 | 5 | 0 | 6 |
| *Romanomermis nielseni* | 6 | 3 | 5 | 0 | 5 |
| *Strelkovimermis spiculatus* | 4 | 3 | 5 | 0 | 6 |
| *Thaumamermis cosgrovei* | 5 | 3 | 5 | 0 | 7 |

There are at least two situations reported in the literature that might be possible alternative solutions to the inconsistencies apparent in the tRNA alignments of the Dorylaimia (i.e. alternative hypotheses). First, throughout the Pulmonata (Mollusca), the tRNA-gly gene has few canonical base-pairs in the Acceptor stem and poor phylogenetic motifs in the 3' part of the stem but not in the 5' part [55]. This can be attributed to an overlap of 4–7 nt with the downstream tRNA-his (or tRNA-cys in one case), which coincides exactly with the poorly conserved part of the tRNA-gly gene. However, this possible explanation does not apply in the case of the Dorylaimia, because the lack of phylogenetic motifs occurs in both the 3' and 5' parts of the Acceptor stem, and there are no suitable gene overlaps involved.

Second, in a somewhat similar set of circumstances Masta [36] and Masta and Boore [17, 18] identified a number of tRNA genes in the Arachnida (Arthropoda; Chelicerata) that are truncated at their 3' end, sometimes including the T arm as well as the Acceptor stem. As with the case of the Pulmonata, post-transcriptional editing was proposed as the mechanism for correcting the affected tRNAs, with the 5' stem acting as a template for the 3' sequence. Unfortunately, application of their approach did not provide a solution to the inconsistencies in the Dorylaimia, once again because the lack of phylogenetic motifs occurs in both the 3' and 5' parts of the Acceptor stem. Conversely, my hypothesis of D- and T-armless tRNAs did not provide a solution to the situation described in the Arachnida.

If my hypothesis for the Dorylaimia proves to be correct, then nematode mitochondrial tRNAs are more bizarre than has been thought. Presumably the situation results from a continuation of the same selective pressure that has resulted in the loss of the D or T arm alone, in the same way that the absence of tRNA genes (e.g. anthozoans, protozoans, fungi, plants; Salinas et al. 2008) can be seen as part of the ongoing process of mitogenomic miniaturization. This would make studying these nematode genomes important for understanding evolutionary processes, as well as for understanding the components of the minimal necessary tRNA [56]. Experimental testing of this hypothesis, by examining the processed tRNA transcripts, would thus be of great value.

## Conclusions

A full comparative analysis, using a multiple alignment of all closely related sequences available, is the most powerful form of annotation procedure for mitochondrial genomes. No computer program currently uses this strategy. For mitochondrial tRNAs there is thus no single program that can be relied upon to produce potential sequences for inclusion in the each gene alignment, as both ARWEN and DOGMA are prone to both false positives and false negatives. Both programs should thus be tried for annotating the full complement of tRNAs in a mitogenome, and this will usually be a successful strategy because their results are often complementary.

The advantage of the approach advocated here is that a coherent comparative framework is applied across all of the alignments, as opposed to the individual strategies used by the original authors. This means that consistent decisions are made to resolve the apparent discrepancies highlighted by the comparative analysis. The results showed that a consistent explanation can, indeed, be provided for most of the discrepancies, across all of the taxonomic groups. It is this very consistency that provides the strongest support for the reliability of the conclusions.

If many tRNA genes are missing from a genome then it is likely that they truly missing, as in some anthozoa, protozoa, fungi and plants. However, if a small number of tRNA genes is apparently missing from a Metazoan mitochondrial genome whose closest relatives all have a full complement of 22 tRNAs (or different tRNAs are missing from different species), then it is likely that they are bioinformatically missing rather than biologically missing. A more thorough bioinformatic analysis may very well reveal them, because absence of evidence is not evidence of absence (or, if you prefer a more poetic version: “no one has the right to say that no water-babies exist, till they have seen no water-babies existing; which is quite a different thing, mind, from not seeing water-babies; and a thing which nobody did, or perhaps ever will do.” Charles Kingsley, *The Water-Babies*, 1863).

## Appendices

Appendix 1. Data sets for each of the alignments
Appendix 2. Changed annotations for each of the genomes
Appendix 3. Alignments for selected genes for selected species from the comparative analyses
Appendix 4. Re-annotation of the *Epiperipatus biolleyi* mitogenome
Appendix 5. Alignments for those genes without D or T arms
Appendix 6. Stem pairing for each of the *Dorylaimia* tRNA gene alignments

Available at (928 KB):
http://acacia.atspace.eu/papers/Supplementary.zip